**Potential tuning in the S-W system. (i) Bringing $T_{c,2}$ to ambient pressure, and (ii) colliding $T_{c,2}$ with the liquid-vapor spinodal.**

C. Austen Angell* and Vitaliy Kapko
School of Molecular Sciences, Arizona State University, Tempe, AZ 8527-1604

**Abstract**
Following Vasisht et al's identification of the second critical point ($T_{c2}$,$P_{c2}$) for liquid silicon in the Stillinger-Weber (S-W) model for silicon, we study the variation of $T_{c2}$,$P_{c2}$ with tetrahedral repulsion parameter in an extension of the earlier "potential tuning" study of this system. We use the simple isochore crossing approach to identify the location of the second critical point (before any crystallization can occur) as a function of the "tuning" or "tetrahedrality", parameter $\lambda$, and identify two phenomena of high interest content. The first is that the second critical point pressure $P_{c2}$, becomes less negative as $\lambda$ decreases from the silicon value (meaning the drive to high tetrahedrality is decreased) and reaches zero pressure at the same value of lambda found to mark the onset of glassforming ability in an earlier study of this tunable system. The second is that, as the $T_{c,2}$ approaches the temperature of the liquid-gas spinodal, $\lambda > 22$, the behavior of the temperature of maximum density (TMD) switches from the behavior seen in most current water pair potential models (locus of TMDs has a maximum), to the behavior seen in empirical engineering multiparameter equations of state (EoS) (and also by two parameter Speedy isothermal expansion EoS) for water, according to which the locus of TMDs of HDL phase has no maximum, and the EoS for HDL has no second critical point. At $\lambda = 23$ the behavior is isomorphic with that of the mW model of water, which is now seen to conform, at least closely, to the "critical point free" scenario for water.

**A. Introduction**

Potential tuning is a strategy for investigating physical properties of a model system as a function of a single parameter which effectively changes the properties of the system systematically between different substances[1]. In the present case the reference substance is elemental silicon, as originally simulated by the Stillinger-Weber potential[2]. These authors had cleverly combined a Lennard-Jones-like pair potential with a 3-body potential that raised the system energy whenever three nearest neighbors subtended an angle different from the tetrahedral angle, 109.47º. The strength of the three-body interaction was controlled by the parameter $\lambda$ in the total interaction potential

$$\upsilon = \upsilon_2(r) + \lambda\, \upsilon_3(r,\theta)$$

In earlier papers[1,3], we had recognized a similarity between the internal action of $\lambda$ on the system and the effect of external pressure, each of which lowers the chemical potential of the liquid relative to the stable (diamond cubic) crystal state, hence lowers the melting point. The object was to seek conditions for slow crystallization, hence glassforming properties, on the computation time scale in the first place[1] and on laboratory time scales in the second[3] . Indeed, it was found that when λ was decreased to below the value 20.5, the original phenomenon of interest, an apparent liquid-liquid[4] (or liquid-amorphous[5]) transition disappeared and glasses were easily formed on reasonable, rather than only quasi-infinite[5], cooling rates. The liquid-amorphous transition had been predicted, on the basis of thermodynamic data, to occur in laboratory silicon by deNeufville and Turnbull[6], and was then identified in different laboratory studies and simulation studies as a transient phenomenon... an Ostwald step on the way to crystallization[7]. Here we will show that this point is of much greater interest than the mere enabling of glass formation.

A related study with features in common but involving the weighting of a component of the potential rather than the character of the potential itself, had earlier been carried out by Lynden-Bell and Debenedetti[8] who weighted the van der Waals component of the SPC-E water potential until it dominated so as to become effectively a Lennard-Jones potential.

More recently, potential tuning has been used to move between silicon and a monatomic water model[9] and, in a rather different case, between Lennard-Jones liquid and a van der Waals ellipsoidal liquid with greatly reduced crystallization propensity, in search of the elusive "ideal glassformer"[10]. The most recent example is that of a simple ionic model of silica in which the charges on silicon and oxygen ions are varied by a single multiplicative factor from the original +4 and -2 values of the so-called WAC model (Woodcock -Angell-Cheeseman[11]) so as to generate a second critical point free from interference by crystallization[12]. With even larger changes the critical point moves below the limit of stability to cavitation (liquid-gas spinodal) and the system line of density maxima moves continuously to higher temperatures, as in the engineering equations of state for water[13,14].

In the present work, we revisit the S-W potential to extend the recent work of Vasisht et al.[15] who studied the λ = 21 (silicon) model at negative pressures to identify the presence of a critical point (CP) where the difference in volume and enthalpy between high and low density phases of liquid silicon, disappear. These workers also mapped out the position of the liquid-vapor spinodal curve that emanates from the *normal* liquid-gas CP, and showed how the temperature of the liquid density maximum narrowly avoids a collision with it.  Our interest lies in finding what happens when the second critical point , which evidently moves to more  negative pressures as λ increases, actually collides with the normal liquid gas spinodal.

In this study we make extensive use of the "isochore-crossing" criterion for the presence of a critical point. This indicator of the presence of a critical point and a liquid-liquid equilibrium domain (or the equivalent unstable van der Waals loop in the case of systems too small to sustain a two phase equilibrium, was pioneered by Skibinsky et al.[16] in their early study of liquid-liquid critical points in hard shoulder pair potential systems of the Hemmer-Stell type[17]. The advantage of this indicator is that it points to criticality before sufficient of the low temperature (usually more crystal-like phase) to promote crystallization, has been produced i.e. it indicates *incipience* of criticality even when longer runs at lower temperatures, where a second phase exists, will produce crystals on short time scales.

The place of lattice models[18] in the understanding of this phenomenology is important but is beyond the scope of the present paper.

## B. New Simulations
### 1. Validation of the Vasisht Tc,Pc for S-W silicon from isochore crossing studies.

Two separate studies of the isochore shapes and relationships for S-W liquid silicon ($\lambda$ = 21) were carried out using the open access LAMMPS software[19]. We ran, simultaneously, 5 different starting densities for the same number of particles, (1718, not 1728[20]), starting at high temperatures where equilibration is very rapid. At each successive 50K lower temperature, the pressure was determined as the average over a 2ns run, starting with the last configuration of the previous run. This corresponds to a cooling rate of 2.5 x $10^{10}$ K/s, some two orders of magnitude slower than those used in now-classical simulations of other glassforming liquids e.g. the archetypal glassformer, silica[21], but faster than that employed for the study of the close cousin of this work, mW water[9]. Examples are shown for two of the chosen densities. The second one is selected because it is the single P-T run, out of two sets of five, in which something unusual occurred.

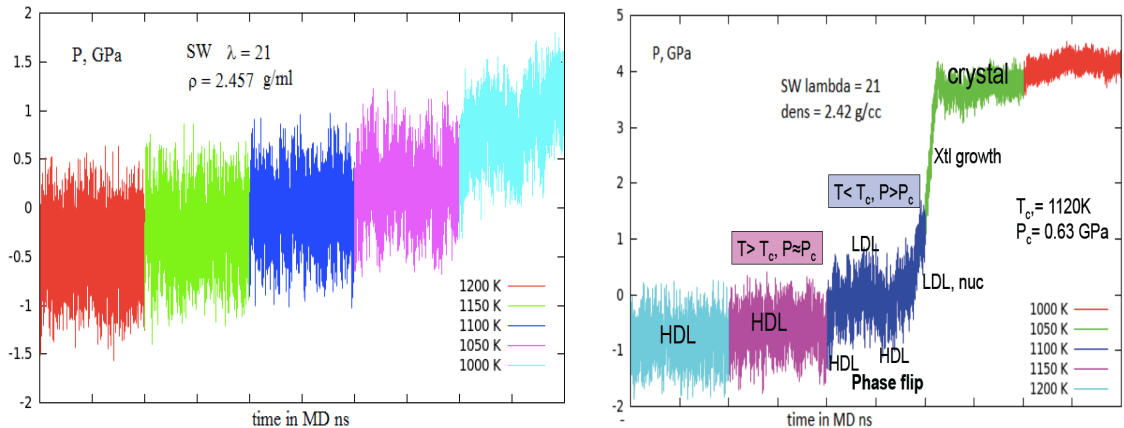

**Figure 1(a)** The last 5 in the sequence of 2ns isothermal holds during the determination of a single isochore (at 2.457 gml$^{-1}$) in the present simulations. for $\lambda$ = 21, the silicon value.
**(b)** Low temperature segment of the run at density of 2.42 gml$^{-1}$ in which a quite sudden event, shown below to be a crystallization, occurred. Structural data for the (hot) crystal phase will be shown later.
================================================================

The individual P,T points extracted from the different segments of the Figure 1 (a) diagram for five different densities for the first of the two λ = 21 studies are shown in Fig. 2(a). Figure 2(a) shows the minimum that is characteristic of water-like liquids that show a density maximum in isobaric cooling experiments, but also shows three of the isochores crossing at a temperature of ~1120ºC and pressure of -0.5 GPa.  This is in good accord with the findings of Vasisht et al. [15]

The crossing point is marked with a red filled circle and the liquid-liquid coexistence line that should exist for samples large enough to support the two coexisting liquid phases in the absence of crystallization, is shown as a red line. Figure 2 (b) shows a second and more tightly packed set of isochores made up from two sets of parallel cooling runs at different densities to confirm the observation of Figure 2 (a). Figure 2(b) however has one additional feature, namely, a sudden jump in pressure for an isochore of density 2.427 gml$^{-1}$ which lies just above the critical density range of 2.35-2.4 where the crossing of the isochores is seen most clearly. This is the isochore illustrated in Figure 1 (b) to which we will return in the discussion section. The fact that only one of the isochores shows this interruption of the otherwise smoothly accelerating increase of pressure, suggests that crystallization of SW21 is not a common event under these isochoric cooling conditions. A little additional cooling will result in the system falling out of equilibrium and entering the glassy state as following figures will show.

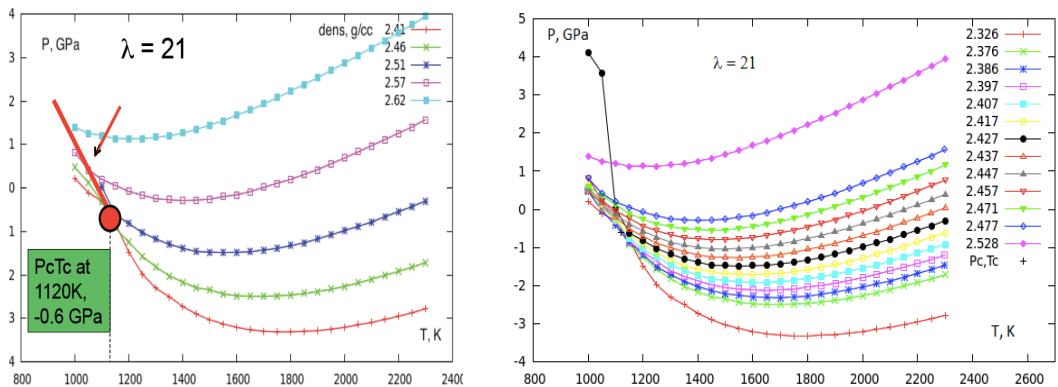

**Figure 2(a).** First set of isochores for the λ=21 case, indicating, by the crossing criterion, the presence of a LL critical point that accords with the findings of Vasisht et al.[15] Note the close similarity to the corresponding set of isochores identifying a LL critical point in the recent charge-modified WAC model of Lascaris[12]. **(b)** a more closely spaced data set, confirming the assignment of Figure 2(a) but also showing the occurrence, at a density just greater than that of the critical density, of a less common event for this relatively slow effective cooling rate, which we identify as a crystallization to the stable (diamond cubic) crystal phase.

## 2. Confirmation of the suspicion that the domain of L-L transitions in the potential-tuned S-W model, ends at a critical point when p = 0.

Armed with this confirmation of the critical isochore behavior we turn to observations made in an earlier study, but left undiscussed. Published only in the SI of ref.[1] was a collection of enthalpy data for the isobaric cooling of samples of λ values which we now reproduce in Figure 3(a). The derivatives of these plots, namely the heat capacity data, are shown in Figure 3(b). It is seen that, as λ decreases towards the value, 20.25, characterizing the beginning of the glassforming range of ref[1], the enthalpy change in the liquid-liquid transition is vanishing and the heat capacity is tending towards a critical-like divergence. To confirm the suspicion that there is indeed a critical point at zero pressure coinciding with the onset of glassforming range we turn to the isochore crossing test for this value of lambda, 20.25. A family of isochores for different densities is plotted in Figure 4, and the crossing at a temperature of ~700K and a pressure of ~0 GPa is clearly indicated. Furthermore we note that at lower temperatures there is a flattening of the isochores which, as a study of mean square displacements shows, is due to the diffusivity falling to very low values with consequent breaking of ergodicity. The critical point and glass transition temperature are almost coincident !! We have seen in other recent studies[22] how dramatically the diffusivity slows down in the vicinity of a liquid-liquid critical point, a feature which distinguishes it from a consolute temperature in a binary solution where only the fluidity and mutual diffusion tend to zero as concentration fluctuations diverge [23].

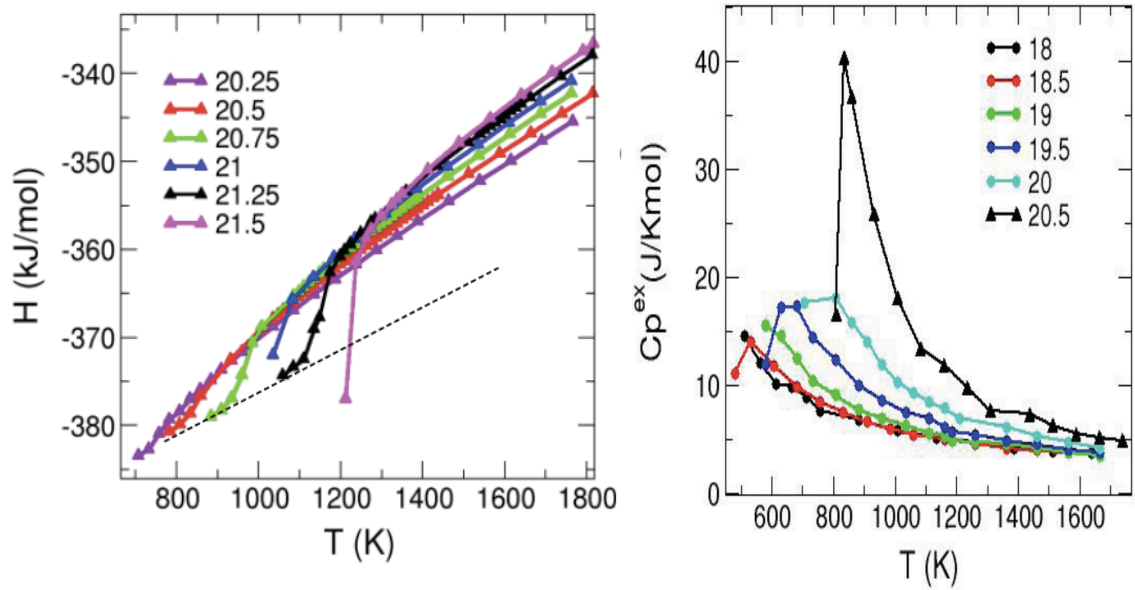

**Figure 3(a)** Variation in liquid enthalpy with temperature at zero pressure, showing also changing value of ΔH, the difference in enthalpy between high temperature and low temperature liquid phases for different λ values. The difference is disappearing as λ approaches 20.25
**(b)** The heat capacity of variation of liquids with temperature as the λ value is tuned down. For λ values below 20.5 the increase is typical of fragile glassformers, being cut off by ergodicity breaking (glass transition)
==========================================================

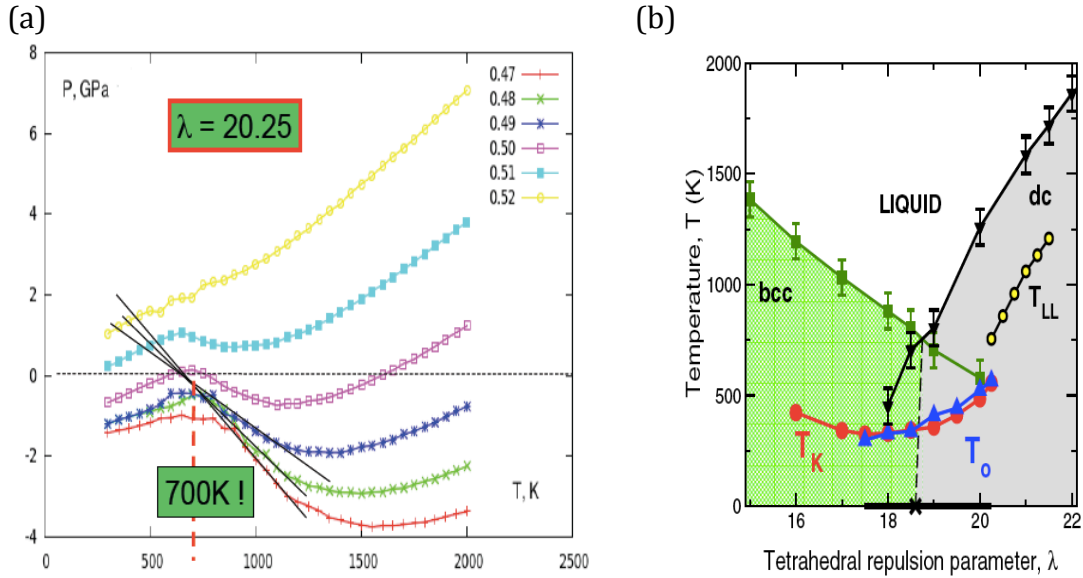

**Figure 4. (a)** A limited set of isochores in the S-W system at λ =20.25, indicating, by crossing, an incipient critical point at 700K, as suggested by Figure 3.
**(b)** the temperature vs tuning potential phase diagram, (from ref. [1] , reproduced by permission of Amer. Inst. Phys.)
========================================================

From this finding it is clear that, in the modified S-W potential model with λ values below 21, the existence of liquid-liquid transitions is terminated by the occurrence of critical points which move from ∼ 750K, close to $T_g$, at zero pressure (for λ = 20.25) to large negative values as λ increases to 21. This trend is leading the liquid-liquid critical point on a collision course with the liquid-vapor spinodal, at which point, new phenomena should emerge. The exploration of phenomenology of modified S-W liquids at λ larger than 21 is the subject of the second part of this paper.

## 3. The S-W model at λ > 21: a collision of $T_{c(L-L)}$ with $T_{s\,(L-V)}$ and consequences.

In this section we make larger changes in the values of λ to ensure that the change in $T_{c\,(L-L)}$ in the λ-tuned S-W system will be large enough to bridge the gap between $T_{c\,(L-L)}$ for the silicon-like potential studied by Vasisht et al.[15] and the limits of stability against cavitation for the liquids (that are expected to lie close to that identified in the latter study). Thus we study the behavior of the isochores at λ values of 22 and 23.

In Figure 5 (a,b) we show the isochores obtained for the λ values of 22 and 23. They have a lot in common, except for the pressure range in which we find them. In neither case is there any tendency of isochores to cross in the way they have in the case of lower λ values. Also in contrast to all except the single isochore that showed a large jump attributed to crystallization, there are now a number that exhibit a smaller but still quite abrupt increases that contrast with the more slowly varying behavior of the majority. The largest such jump is that at density 2.427 gml$^{-1}$, which is highlighted by a box in Figure 5(a) for the λ = 22 case. We note that it jumps right across the more smoothly varying isochore of higher density (2.477 gml$^{-1}$). To see what might be responsible, liquid-liquid transition or crystallization, we plot the detailed sequence of isothermal holds in Figure 6a, for the λ = 22 case, and the radial distribution function for the 1200 K structure in Figure 6b.

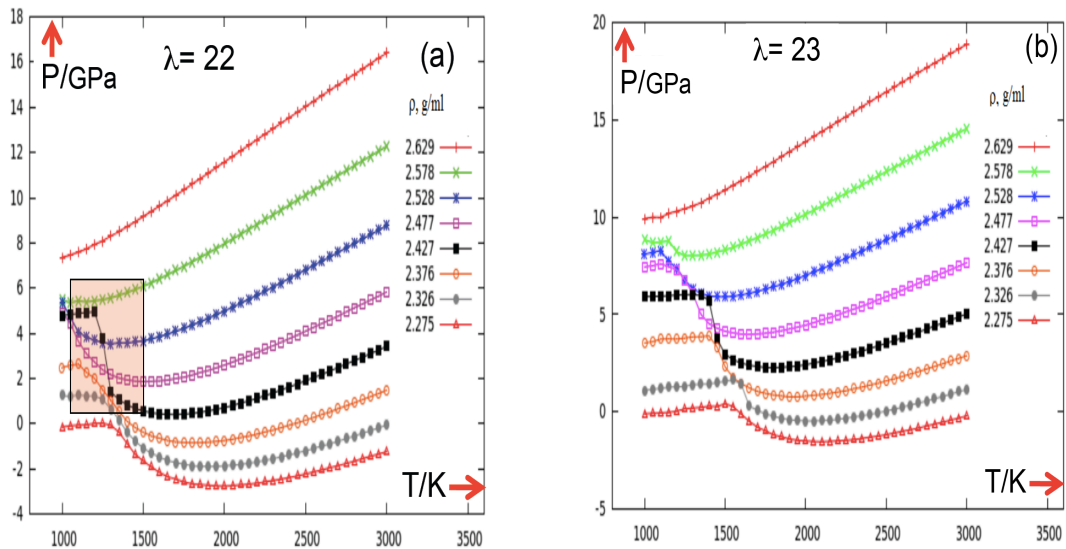

**Figure 5.** Isochores at the higher λ values of 22 and 23. Note the jumps of pressure for the 2.427 isochores and the smaller jump at 2.326 in the case of λ = 23. Note also the plateauing of the isochore immediately after the jump, indicative of solid behavior. Apart from the false crossings caused by the jumps, there are no isochore crossings of the type seen in earlier isochore sets. Note that entire isochore set is displaced to higher P for λ=23.
============================================================

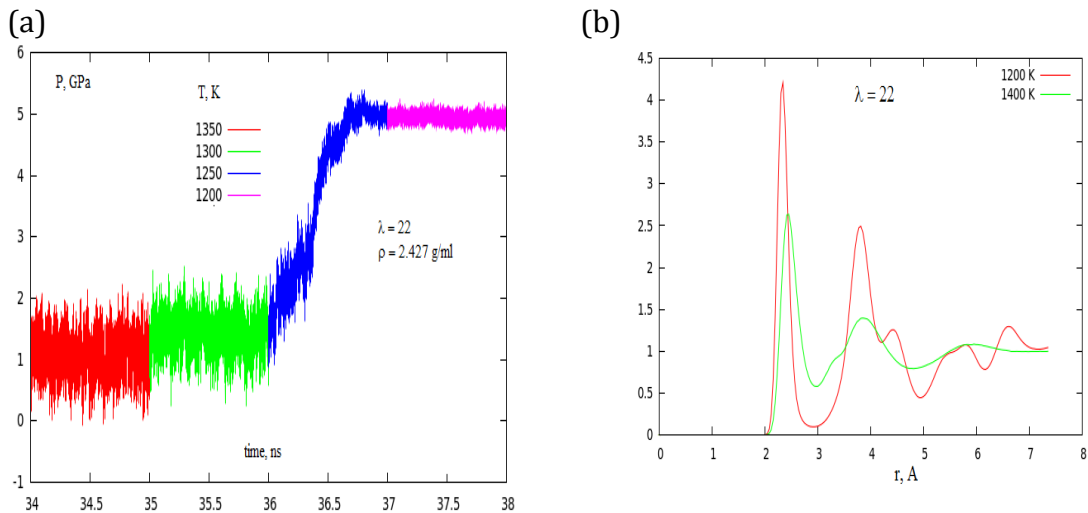

**Figure 6.** (a) Details of the pressure jump in the 2.427 gml$^{-1}$ isochore between 1300 and 1200K for $\lambda$ = 22 (b) the radial distribution function for the 1200 K structure, showing that the jump is due to crystallization. An implication is that the other isochoric coolings in Figure 5a gave glasses. The smaller jump in pressure for the isochore at density 2.326 gml$^{-1}$, when $\lambda$ =23, also is associated with crystallization, giving the same change in RDF seen in Figure 6(b). The thermal broadening of the crystalline RDF vanishes on reduction of temperature to ambient.

=========================================================

To confirm that the jump in isochore pressure is indeed due to crystallization we plot the mean square displacements for the $\lambda$ =23, $\rho$ = 2.427 gml$^{-1}$ case for temperatures from 1600K down to 1400 K. We observe that the MSD drops to zero on the scale of the plot between 1450 and 1400K, consistent with a liquid-to-crystal phase transition, and that this is the temperature range of the sudden pressure jump in Figure 5b (see box). At 1450K the slope of the plot corresponds to a diffusivity of order 10$^4$ cm$^2$s$^{-1}$, considerably more fluid than water (but consistent with liquid metal character).

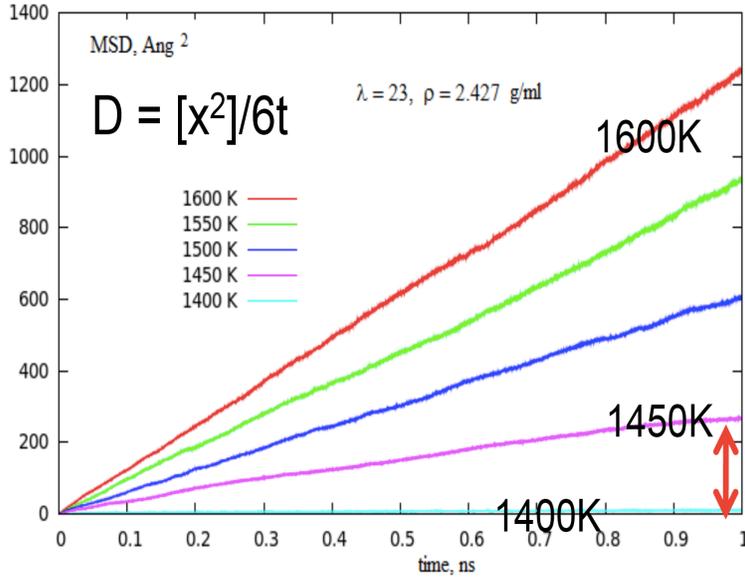

**Figure 7**. Mean square displacements vs time for the l = 23 system for a series of temperatures including the values 1450 and 1400 K on either side of the isochore pressure jump. The cessation of diffusion on this time scale supports the disappearance of the liquid phase in this temperature interval i.e. across the pressure jump.

==========================================================

Finally, we need to confirm that this reversal in behavior is associated with the changes in relative positions of the density maximum and the liquid-vapor spinodal. The liquid-vapor spinodal is revealed approximately by observing the pressure during isothermal (stepwise) expansion. The pressure will suddenly regress towards zero when a cavity forms at the kinetic stability limit. Exactly where this will be observed of course depends on the rate of expansion and we have used a relatively low rate so as to ensure good pressure data up to the cavitation point. Vasisht et al.[15] established the L-V spinodal for two different expansion rates, for the case of $\lambda$ = 21. Our data for $\lambda$ = 21 are consistent with theirs for the slower rate.

In Figure 8 we show data for $\lambda$ = 23 in which expansions at temperatures 1700K and above are seen to be well behaved, and yield well-defined stability limits. Runs at lower temperatures appear erratic and to have higher tensile limits, but this is because they have crossed the LL line during the expansion and, as in the case of isochores for this value of $\lambda$, have become unstable to crystallization.

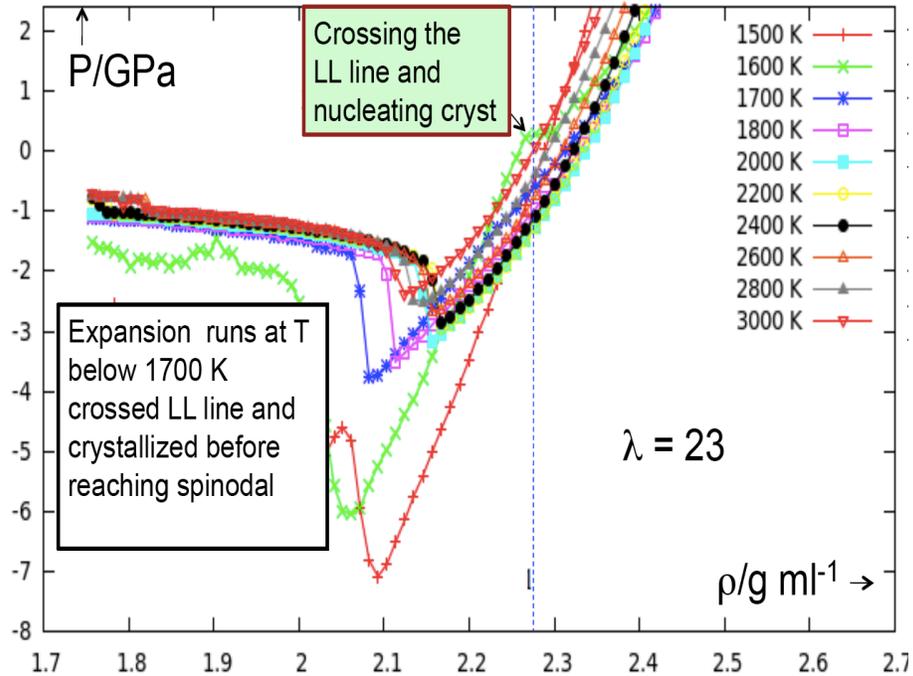

**Figure 8.** Determination of the limit of liquid stability against isothermal volume increase for the λ = 23 system. This is a cavitation limit which is kinetically controlled by vacuum bubble nucleation. The thermodynamic limit (spinodal) lies at slightly more negative pressure where $(\partial\rho/\partial_T$ ---> $\infty$. Note that the two lowest temperature isotherms at 1500 and 1600K show different behavior, and were found to be crystalline in consequence of crossing the LL line at which crystallization rates are enhanced by the high density liquid to low density liquid (HDA-LDA) transition. (see also ref. [15])

## Discussion

As a preliminary to our discussion of the significance of these findings, we show, in Figure 9, a schematic of what our data, in combination with those of Vasisht et al.[15], are suggesting. We remember that each value of lambda creates a new and distinct substance with its own liquid gas spinodal, and its own second critical point.

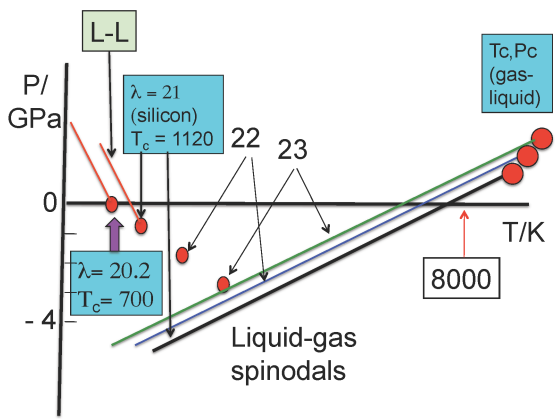

**Figure 9.** Schematic of the behavior of critical points and liquid-gas spinodals identified in the work of Vasisht et al. and the present work at lower and higher values of the tuning parameter λ. Figure indicates the collision course of the liquid-liquid critical point with the liquid gas spinodal, as λ is increased towards the value 23.

We have isochore crossing evidence for second critical points at λ = 20.2 and 21.0 which establish a negative slope for CP in the p,T plane. The points at λ = 22 and 23 cannot be assigned on the basis of isochore crossings, perhaps because the isochores were not obtained at low enough densities at least in the case of λ = 22. Rather their positions are based on the behavior of the isochore minima (density maxima), and the reversal of the pressure at which the minimum pressure at a given density is reached (seen in Figure 5(a) and (b)).

Figure 10 then shows the combination of Figure 8 data on the spinodal limit to cavitation, with the isochores that we have already presented for this lambda value in Figure 5b.  Figure 10a has much in common with Figure 10b which is the display of isochores, and the spinodal limits they imply (by the criterion of diverging compressibility) for pure water based on the Haar-Gallagher-Kerr equation of state (EoS)[24]. For many years this EoS, based on precise measurements made on the stable states of pure water, was used as the best available representation of PVT properties of this substance (the National Bureau of Standards equation)

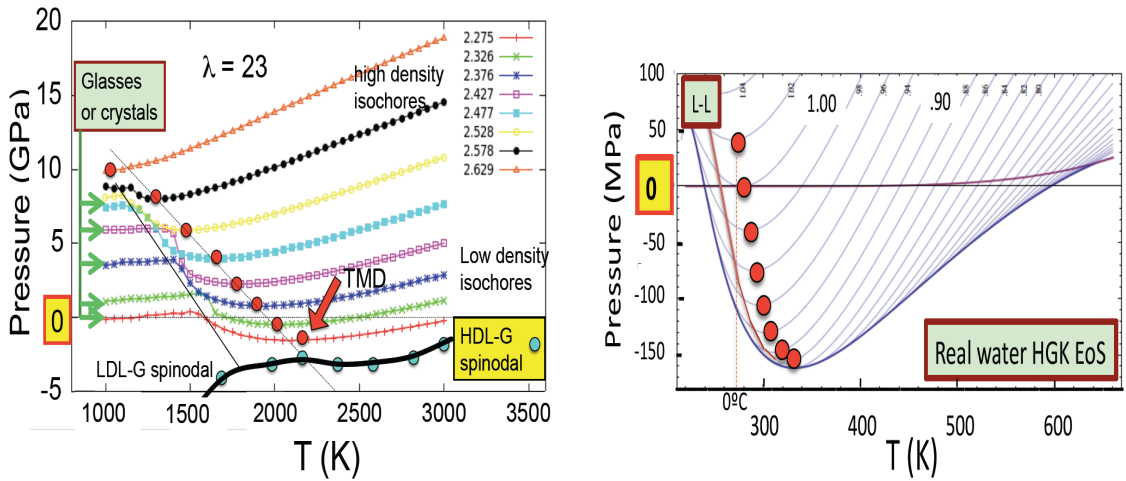

**Figure 10**  (a) Combination of isochoric P-T data, obtained during cooling of S-W-23, with data on the stability limit against cavitation during slow isothermal expansion of the same system. The corresponding L-G spinodal for λ = 21 lies at slightly more negative pressures.   (b) Isochores and their spinodal limits for real water (HDL), according to the HGK equation of state. Temperatures of maximum density (TMD) and their merging with the spinodal, are noted. The possible line of liquid-liquid transitions that is suggested by the behavior of S-W models of lower lambda values, is indicated by the line between the locus of TMD and the HGK low temperature spinodal limit for HDL.
=============================================================

Of course the extrapolations of such a multiparameter EoS into P,T regions outside the range of the data used to establish the equation parameters, in particular into the negative pressure domain, are in principle quite unreliable.

However, it has been impressive how accurately the HGK equation was (i) able to predict the existence of a domain of diverging compressibilities in the supercooled temperature range for water at positive pressures[25,26] and (ii) able to predict the existence of a limit to the range of densities of water that would cavitate under isochoric cooling protocols that obliged them to enter the negative pressure domain[27-30], (isochores of density 1.0 -0.91 gml$^{-1}$). Now we see qualitatively similar behavior in a model system that is derived, by potential tuning, of a much-studied pair potential model for silicon. Indeed, the successful mW model of water that has received much attention since the seminal paper of Molinero and Moore[9] is essentially a rescaled version of the S-W model with the choice $\lambda$ = 23 whose behavior in the negative pressure domain has been displayed in Figures 5 and 7-9.

The HGK equation, being based on the behavior of water in its stable liquid states, of course cannot predict the occurrence of either crystallization or liquid-liquid phase transitions. Such excursions to *alternative* free energy surfaces require nucleation and growth of the second phase[31].

The inability of mW water to form a separate liquid phase before nucleation of the stable ice phase during isobaric cooling, has been studied in great detail by Molinero and coworkers[9,32,33] and also by Limmer and Chandler[34]. However, by use of a cooling rate similar to that of the present study , Molinero and Moore[9] reported suppression of crystals and formation of a glassy phase that could be further relaxed without interference from crystalline ice.

Also, the behavior of mW water during isochoric cooling has not been established. In real water it has long been known that the anomalous fluctuations in which ice nuclei form[32] are damped out almost completely during isochoric cooling, so the probability of crystallization will be reduced, also by the concomitant increases in pressure that must occur. We also note the new window being opened on the water problem using the supercooling abilities of newly available ideal solutions with organic and inorganic "ionic liquids" as second components[35] that can serve as proxies for pressure increases on the solute-free liquid.

While the line of LL transitions shown in Figure 9b is rendered unphysical by fast crystallization in the SW-23 model and its water analog, the situation with real water at negative pressure is less clear.  This is because the intrinsic mobilities in water are much smaller than in atomic models due to hydrogen bonds and their lifetimes,  and also because of the major kinetic slowdowns expected, from all models of tetrahedral liquids, as the tetrahedral water network is stabilized at negative pressure. Indeed Pallares et al.[36] have given Brillouin scattering evidence that the anomalous zone can be passed through continuously during isochoric cooling in the negative pressure domain, (though the exact origin of the anomaly is not clear from their work[37]).

Some additional insights into the behavior of tetrahedral network systems in the absence of crystallization may be obtained from the recent study of the WAC ionic model of silica under charge tuning, recently reported by Lascaris[12]. This study also employed the isochore-crossing criterion to identify when true criticality could be obtained,d and showed how, at extreme tuning, the critical point could be pushed into the liquid-gas spinodal in association with a flattening of the TMD locus, as seen in the HGK equation[24]-based findings of Figure 9b and less clearly in figure 9a for the present study.

**Concluding remarks.**

Potential tuning MD is a valuable tool in the search for unusual phenomena in condensed matter physics. It is hoped to extend the present approach to the tuning of the WAC model of silica, i.e. by adding a variable three body repulsive contribution to the total energy to enhance its tetrahedrality and thereby reach critical conditions by a route alternative to that of Lascaris who succeeded by tuning only the two body interactions. Regardless of how the tuning is enabled it seems that, at least in tetrahedral systems, the direction of tuning needed to achieve criticality is also the direction of lower pressure, and that in many cases, the critical point is only encountered in the negative pressure domain.

In the special case of water, a very successful model is the mW model which is a scaled-down version of the present model with $\lambda = 23$. We have shown in the present study that this model has no condition in which the isochores tend to cross on decreasing temperature, i.e. even if the LDL phase could be protected against crystallization, the model would lack an HDL-LDL critical point because the critical point has been pushed into the gas-liquid spinodal. mW water, in other words, seems to be an example (or a close approach to one) of the "critical-point-free" scenario proposed for real water[38] and earlier suggested by all the engineering equations of state (HGK and four others displayed in the major work on IAP-WS95[13]). It is notable that, in a new ab initio study of liquid silicon itself, Zhao et al.[39] find the same scenario applying to their findings.

**Acknowledgements**
This work was supported by the NSF under collaborative grant CHE12-13265

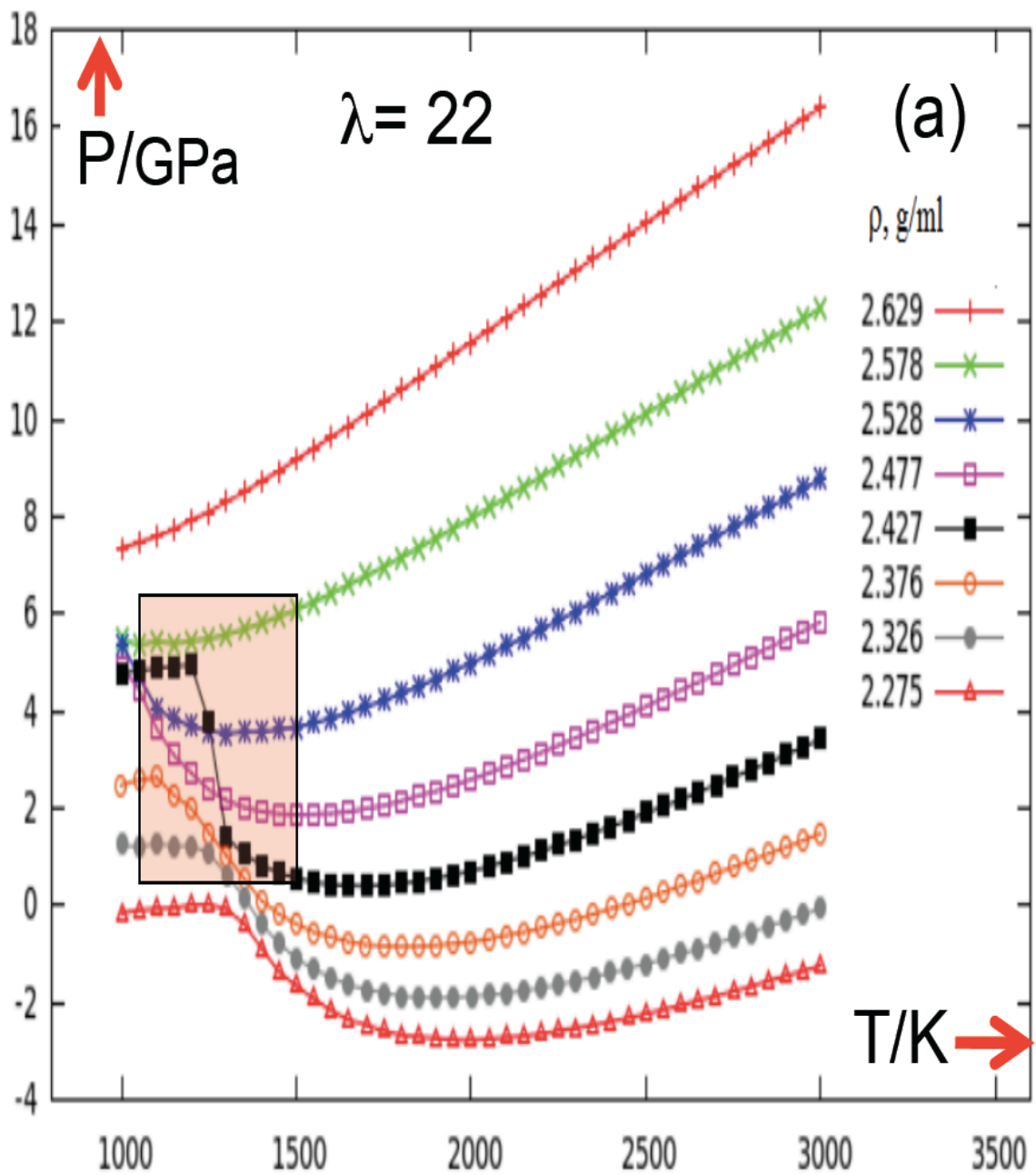